\documentclass[fleqn,10pt]{wlscirep}
\usepackage[utf8]{inputenc}
\usepackage[T1]{fontenc}

\usepackage{jabbrv}
\usepackage{empheq}
\usepackage{colortbl}
\definecolor{VeryLightBlue}{rgb}{0.94, 0.97, 1.0}
\definecolor{LightBlue}{rgb}{0.68, 0.85, 0.9}

\title{Learning the Stellar Structure Equations via Self-supervised Physics-Informed Neural Networks}

\author[1,+,*]{Manuel Ballester}
\author[2,+]{Santiago Lopez-Tapia}
\author[1, 3, 4]{Seth Gossage}
\author[1,2]{Patrick Koller}
\author[2]{Philipp M. Srivastava}
\author[2]{Ugur Demir}
\author[1]{Yongseok Jo}
\author[5]{Almudena P. Marquez}
\author[6]{Christoph Wuersch}
\author[7]{Souvik Chakraborty}
\author[1,3,4]{Vicky Kalogera}
\author[1,2]{Aggelos Katsaggelos}

\affil[1]{SkAI Institute (NSF–Simons AI Institute for the Sky), Chicago, IL, USA}
\affil[2]{Department of Electrical and Computer Engineering, Northwestern University, Chicago, IL, USA}
\affil[3]{CIERA, Northwestern University, Chicago, IL, USA}
\affil[4]{Department of Physics and Astronomy, Northwestern University, Chicago, IL, USA}
\affil[5]{Department of Mathematics, University of Cadiz, Cadiz, Spain}
\affil[6]{OST Eastern Switzerland University of Applied Sciences, Switzerland}
\affil[7]{Indian Institute of Technology (IIT) Delhi, New Delhi, India}

\affil[+]{These authors contributed equally to this work}
\affil[*]{Corresponding author: manuel.ballester@northwestern.edu}


\begin{abstract}
Stellar astrophysics relies critically on accurate descriptions of the physical conditions inside stars. Traditional solvers such as \texttt{MESA} (Modules for Experiments in Stellar Astrophysics), which employ adaptive finite-difference methods, can become computationally expensive and challenging to scale for large stellar population synthesis ($>10^9$ stars). In this work, we present an self-supervised physics-informed neural network (PINN) framework that provides a mesh-free and fully differentiable approach to solving the stellar structure equations under hydrostatic and thermal equilibrium. The model takes as input the stellar boundary conditions (at the center and surface) together with the chemical composition, and learns continuous radial profiles for mass $M_r(r)$, pressure $P(r)$, density $\rho(r)$, temperature $T(r)$, and luminosity $L_r(r)$ by enforcing the governing structure equations through physics-based loss terms. To incorporate realistic microphysics, we introduce auxiliary neural networks that approximate the equation of state and opacity tables as smooth, differentiable functions of the local thermodynamic state. These surrogates replace traditional tabulated inputs and enable end-to-end training. Once trained for a given star, the model produces continuous solutions across the entire radial domain without requiring discretization or interpolation. Validation against benchmark \texttt{MESA} models across a range of stellar masses yields a Mean Relative Absolute Error of $3.06\%$ and an average $R^2$ score of $99.98\%$. To our knowledge, this is the first demonstration that the stellar structure equations can be solved in a fully self-supervised and data-free fashion employing PINNs. This work establishes a foundation for scalable, physics-informed emulation of stellar interiors and opens the door to future extensions toward time-dependent stellar evolution.
\end{abstract}
\begin{document}

\maketitle

\section{Introduction}
Understanding the internal structure of stars remains one of the central problems in stellar astrophysics \cite{kippenhahn1990stellar, hansen2004overview}. The internal radial profiles of fundamental physical quantities (such as pressure, density, temperature, luminosity and enclosed mass) govern the observable properties of the star, including its total luminosity (magnitude), effective temperature (color), total radius, and nucleosynthetic outputs \cite{prialnik2009introduction, clayton1983principles}. Accurately modeling these internal structures is therefore essential in order to connect theoretical predictions with observations across a wide range of astrophysical phenomena.

The open-source code \texttt{MESA} (Modules for Experiments in Stellar Astrophysics) \cite{paxton2011modules, paxton2013modules, paxton2015modules, paxton2018modules, paxton2019modules} represents the state-of-the-art in stellar structure modeling, combining macroscopic conservation laws with detailed microphysical processes (such as opacity, equation of state, and nuclear reaction networks). Despite its accuracy and flexibility, \texttt{MESA} remains computationally intensive for certain applications. Each stellar model typically requires iterative finite-difference solvers, repeated interpolation of large tabulated datasets, and adaptive mesh refinement, resulting in runtimes on the order of hours per star. While this cost is acceptable for individual studies, it becomes prohibitive for large-scale use cases such as stellar population synthesis \cite{bruzual2003stellar, byrne2024bpass, conroy2010propagation, fragos2023posydon, andrews2025posydon}, which may require evaluating billions of single and binary stellar models.

This computational challenge is expected to intensify with the advent of next-generation surveys, such as the Vera C. Rubin Observatory LSST \cite{ivezic2016lsst}, which will produce massive volumes of data on stellar populations. These efforts demand fast, scalable, and physically consistent models capable of evaluating stellar properties across broad ranges of masses and compositions, often in real time. This motivates the development of alternative approaches that retain the physical fidelity of classical solvers while significantly improving computational efficiency.

\begin{figure*}[h!]
\centering
\includegraphics[width=0.9\textwidth]{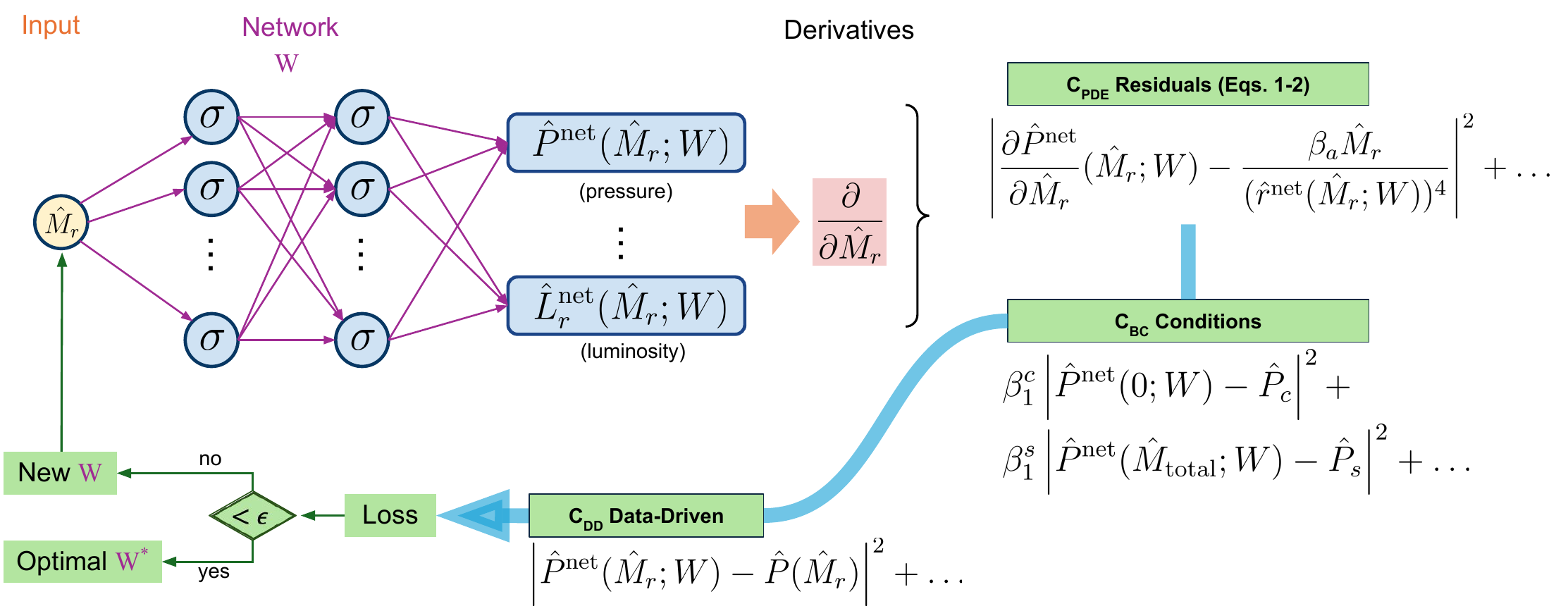}
\caption{Schematic of the Physics-Informed Neural Network (PINN) framework for stellar structure modeling. The network maps the normalized enclosed mass ($\hat{M}_r$) to the stellar state variables (pressure, radius, temperature, and luminosity). The training objective combines a physics-based loss $C_{\mathrm{PDE}}$, defined as the $L^2$ norm of the equation residuals at collocation points, and a boundary-condition term $C_{\mathrm{BC}}$. An optional data-driven term $C_{\mathrm{DD}}$ can be included in supervised settings, but is omitted in the fully self-supervised formulation considered in this work.}
\label{fig:pinn_diagram}
\end{figure*}

Physics-Informed Neural Networks (PINNs) have recently emerged as a promising framework for solving differential equations by embedding physical laws directly into the training process \cite{raissi2019physics, mao2020physics, cai2021physics, karniadakis2021physics, patel2022physics, djeumou2022neural, gazoulis2025stability}. Instead of relying on labeled input-output data, PINNs minimize the residuals of the governing equations, effectively using the equations themselves during training. Through automatic differentiation, the network can represent both the solution and its derivatives, enabling the continuous enforcement of differential constraints across the domain. In addition, architectural design choices can be used to encode known physical structure, further restricting the space of admissible solutions.

Despite their rapid development, the application of PINNs to stellar astrophysics remains largely unexplored. Existing machine learning approaches for stellar modeling typically rely on supervised learning using precomputed stellar models \cite{ness2015cannon, ness2018data, ho2017label, sharma2020application, dafonte2020blended, weaver2000spectral, leung2019deep}. While effective within their training domain, these approaches inherit biases from the underlying simulations and do not generalize reliably beyond them. In contrast, a fully self-supervised PINN trained solely on the governing equations and boundary conditions constitutes a data-free and independent solver \cite{lu2025unsupervised, wan2025physics, jiang2025unsupervised}.

However, directly applying standard PINNs to stellar structure problems proves challenging. The stellar structure equations are highly nonlinear, stiff, and tightly coupled, with solutions spanning many orders of magnitude and requiring strict enforcement of boundary conditions at both the stellar center and surface. In practice, naive PINN formulations often suffer from a number of challenges \cite{bonfanti2024challenges, yan2024improved, jahani2024enhancing, zou2025accelerating, zhao2024comprehensive, chai2024overcoming, mao2023physics, jia2026multi, lu2021physics}, including the slow convergence, poor representation of sharp gradients, and violations of physical constraints, making them insufficient for accurate stellar modeling.

A key contribution of this work is to show that accurate stellar structure modeling with PINNs does not arise from a single architectural choice, but rather from a carefully designed combination of recent advances in physics-informed learning. Our framework integrates: (i) hard-constraint enforcement of boundary conditions through analytic transformations, ensuring exact satisfaction of central and surface conditions \cite{lu2021physics, marquez2017imposing, alkhadhr2023wave, xiao2024hard} ; (ii) auxiliary neural networks with Random Fourier Feature embeddings \cite{rahimi2007random, wu2025iterative, sallam2023use, du2026label, ding2025physics, xiong2025high} to model tabulated microphysics (equation of state and opacity) as smooth, differentiable functions; (iii) a SIREN-based architecture \cite{sitzmann2020implicit, pezzoli2023implicit, zhang2026adaptive} for the main PINN to efficiently capture high-frequency features with a compact model; (iv) the Stochastic Projection PINN (SP-PINN) approach \cite{navaneeth2023stochastic, garg2025neuropinns} for gradient-free approximation of PDE derivatives, reducing computational cost; and (v) an active learning strategy based on Residual-Based Attention \cite{anagnostopoulos2024residual, ramirez2025residual}, which adaptively concentrates collocation points in regions where the solution is most challenging.

While each of these components has been explored in isolation in prior work, their integration and adaptation to the stellar structure problem are essential to overcome the multiscale behavior, stiffness, and strict physical constraints of stellar interiors. Together, they enable stable and accurate training of a fully self-supervised model that acts as a continuous, mesh-free solver. The following sections describe each of these components in detail and how they are combined into a unified framework.

In this work, we develop such an self-supervised PINN framework to directly solve the four canonical stellar structure equations under the assumption of hydrostatic and thermal equilibrium. The network takes as input the independent variable (either the radial coordinate $r$ or, equivalently, the enclosed mass $M_r$) together with the stellar chemical composition $(X, Y, Z)$. The energy generation rate $\epsilon$, including nuclear reactions and neutrino losses, is computed using classical finite-difference microphysical routines from \texttt{MESA}, ensuring physical consistency. The opacity $\kappa$ and the equation of state (EoS) are modeled through auxiliary neural networks trained on tabulated data, enabling end-to-end differentiability.

The resulting model learns continuous radial profiles of the stellar properties (mass enclosed, pressure, density, temperature, and luminosity) at randomly sampled collocation points, producing solutions that can be evaluated at arbitrary locations without discretization or interpolation. This mesh-free and differentiable formulation makes the approach particularly well suited for applications such as sensitivity analysis, inverse problems, and large-scale population synthesis.

Validation against benchmark \texttt{MESA} models across a range of stellar masses demonstrates high accuracy, with a Mean Relative Absolute Error (MRAE) of $3.06\%$ and an average $R^2$ score of $99.98\%$. While the present work focuses on equilibrium stellar structures, we also explore preliminary extensions toward time-dependent evolution, highlighting both the potential and current limitations of the approach.

In summary, we introduce a data-free neural solver for the stellar structure equations that integrates realistic microphysics, enforces boundary conditions exactly, and combines multiple recent advances in physics-informed learning into a unified framework. This work represents a first step toward fully self-supervised, physics-informed modeling of stellar interiors with realistic microphysics, establishing a foundation for scalable simulations and future extensions to time-dependent stellar evolution and more complex astrophysical systems.

\section{Methodology}
\label{sec:methodology}

\subsection{Overview}
\label{sec:overview}
As mentioned above, the goal of this work is to construct a PINN model that learns the internal structure of a star in hydrostatic and thermal equilibrium directly from the governing differential equations, without requiring precomputed training data. In this section, we describe the physical formulation, introduce the main variables and notation, and outline the overall modeling strategy.

A star in equilibrium is characterized by the radial profiles of several coupled physical quantities: the pressure $P$, which balances gravitational contraction; the density $\rho$, which determines the local mass distribution; the temperature $T$, which governs the thermal state and energy transport; and the luminosity $L_r$, which represents the net energy flux passing through a spherical shell at radius $r$. These quantities are related through a system of coupled differential equations known as the stellar structure equations, expressing conservation of mass, momentum (hydrostatic equilibrium), energy, and energy transport \cite{prialnik2009introduction}.

A fifth quantity, the enclosed mass $M_r$, defined as the total mass within radius $r$, plays a central role. Since $M_r$ increases monotonically with $r$, there exists a one-to-one correspondence between these variables. We exploit this property by adopting $M_r$ as the independent variable. This Lagrangian formulation avoids coordinate singularities at the stellar center and leads to improved numerical stability.

The PINN is trained by minimizing the residuals of the stellar structure equations at a set of collocation points sampled across the domain $\hat{M}_r \in [0, \hat{M}_{\mathrm{total}}]$. The network takes the normalized enclosed mass $\hat{M}_r$ as input and predicts the stellar variables $\hat{r}$, $\hat{P}$, $\hat{T}$, and $\hat{L}_r$. Please observe that, in order to improve numerical stability, all physical quantities are normalized using solar reference values (we are using the standard dimensionless notation for the quantities with a hat, such as $\hat{M}_r = M_r/M_\odot$, $\hat{r} = r/R_\odot$, and $\hat{P} = P/P_\odot$). The boundary conditions at the stellar center and surface are enforced in the PINN model using a hard-constraint formulation, in which analytic transformations ensure that the network outputs satisfy the prescribed values by construction.

The system is closed through three microphysical relations: the equation of state (EOS), the opacity $\kappa$, and the energy generation rate $\epsilon$. We adopt a hybrid strategy. The EOS and opacity are modeled using auxiliary neural networks trained on tabulated microphysics, providing smooth and differentiable surrogates. In contrast, the energy generation rate (due to its stiffness and complexity) is computed using established finite-difference based routines from \texttt{MESA}.

\subsection{Stellar Structure Equations}
\label{sec:governing_equations}
Under the assumption of spherical symmetry, the internal structure of a star is described by a system of coupled differential equations governing mass conservation, momentum balance, energy transport, and energy conservation. In their most general form, these equations depend on both the enclosed mass coordinate $M_r$ and time $t$, allowing for stellar evolution. Following the standard formulation, the time-dependent stellar structure equations can be written as
\begin{align}
    \frac{\partial \hat{P}}{\partial \hat{M}_r} &= -\frac{\beta_a \hat{M}_r}{\hat{r}^4}
    - \frac{\beta_e}{\hat{r}^2} \frac{\partial^2 \hat{r}}{\partial t^2},
    \label{eq:hydrostatic_time} \\[6pt]
    \frac{\partial \hat{r}}{\partial \hat{M}_r} &= \frac{\beta_b}{\hat{r}^2 \hat{\rho}},
    \label{eq:mass_time} \\[6pt]
    \frac{\partial \hat{T}}{\partial \hat{M}_r} &= -\frac{\beta_c \hat{L}_r}{\hat{r}^4} \, \nabla(\hat{M}_r,t),
    \label{eq:temperature_time} \\[6pt]
    \frac{\partial \hat{L}_r}{\partial \hat{M}_r} &= \beta_d \left[\epsilon(\hat{M}_r,t) - T \frac{\partial S}{\partial t}\right],
    \label{eq:luminosity_time}
\end{align}
where the additional term in Eq.~\eqref{eq:hydrostatic_time} accounts for dynamical acceleration, and the entropy term in Eq.~\eqref{eq:luminosity_time} captures time-dependent thermal evolution. The dimensionless constants $\beta_a,\beta_b,\beta_c,\beta_d,\beta_e$ absorb physical constants and normalization factors and are further discussed and derived in the Supplementary Material (Section 1).

This system describes the full time evolution of a star, including dynamical adjustments, thermal relaxation, and changes in internal structure driven by nuclear processes and entropy variations. However, solving this fully time-dependent system is significantly more challenging due to stiffness, multi-scale coupling, and the need to track entropy evolution consistently. 

The stellar structure equations are fully specified by the set of boundary conditions and microphysical inputs, primarily determined by the total initial stellar mass $M_T$ and its chemical composition $(X,Y,Z)$, denoting the mass fractions of hydrogen, helium, and heavier elements (metals), with $Y = 1 - X - Z$. These quantities define the global properties of the star and enter the system through the equation of state, opacity, and energy generation rate. In this time-dependent formulation, the composition evolves according to nuclear reaction networks, introducing additional equations of the form $\frac{dX}{dt}$, $\frac{dY}{dt}$, and $\frac{dZ}{dt}$.

In this work, we focus on stars in hydrostatic and thermal equilibrium \cite{pols2011stellar, macdonald2015equations}. Under these assumptions, the time-dependent terms vanish,
\begin{equation}
\frac{\partial^2 \hat{r}}{\partial t^2} \approx 0, 
\qquad
\frac{\partial S}{\partial t} \approx 0,
\qquad
\frac{dX}{dt} \approx \frac{dY}{dt} \approx \frac{dZ}{dt} \approx 0,
\end{equation}
A central quantity in the formulation is the dimensionless temperature gradient,
\begin{equation}
\nabla = \frac{d \ln T}{d \ln P},
\end{equation}
which determines the dominant energy transport mechanism. In stellar interiors, energy is transported by radiative diffusion along or by radiative diffusion and convection together, depending on local stability conditions. This is modeled through the piecewise relation
\begin{equation}
\nabla =
\begin{cases}
\nabla_{\mathrm{rad}}, & \text{if } \nabla_{\mathrm{rad}} \le \nabla_{\mathrm{ad}}, \\[6pt]
\nabla_{\mathrm{conv}}, & \text{if } \nabla_{\mathrm{rad}} > \nabla_{\mathrm{ad}},
\end{cases}
\label{eq:nabla_piecewise}
\end{equation}
where $\nabla_{\mathrm{ad}}$ is the adiabatic gradient obtained from the equation of state.

The radiative gradient is given by
\begin{equation}
\nabla_{\mathrm{rad}} = \frac{3 \kappa \hat{P} \hat{L}_r}{16 \pi a c G \hat{M}_r \hat{T}^4},
\end{equation}
while convective regions are identified through the Schwarzschild criterion $\nabla_{\mathrm{rad}} > \nabla_{\mathrm{ad}}$. In these regions, the effective gradient $\nabla_{\mathrm{conv}}$ is computed using mixing-length theory \cite{paxton2011modules, cox1968principles, henyey1965studies} (further detailed in the Supplementary Material, Section 2), which provides a local approximation to turbulent energy transport and drives the temperature gradient toward the adiabatic limit.

The resulting system defines a nonlinear boundary-value problem with conditions imposed at both the stellar center and surface. At the center ($\hat{M}_r = 0$), regularity requires $\hat{r}(0) = 0$ and $\hat{L}_r(0) = 0$ \cite{pols2011stellar}, while at the surface ($\hat{M}_r = \hat{M}_{\mathrm{total}}$), the stellar variables match the specific atmospheric boundary conditions \cite{hauschildt1999nextgen, hauschildt1999nextgen2, castelli2003modelling, allard2001limiting}. The details about the specific boundary value calculations can be found in \cite{paxton2011modules} and in Section 3 of the Supplementary Material.

While the general time-dependent formulation in Eqs.~\eqref{eq:hydrostatic_time}–\eqref{eq:luminosity_time} provides the modeling of the stellar evolution, the present work mainly focuses on solving the equilibrium system. We will also explore in this manuscript some preliminary extensions of our framework to include time as an input, highlighting both the potential and current limitations of this approach.

\subsection{Microphysical Closures}
\label{sec:microphysics}
The stellar structure equations involve five unknown fields ($\hat{P}$, $\hat{r}$, $\hat{\rho}$, $\hat{T}$, $\hat{L}_r$) but provide only four differential relations. Closing the system requires additional microphysical relations: the equation of state, opacity, and energy generation rate. These quantities are introduced below and treated using a hybrid strategy combining learned surrogates and fixed physics operators.

\subsubsection{Equation of State as an Auxiliary Network}
\label{sec:eos}
The equation of state provides the thermodynamic closure relating pressure, density, temperature, and composition. While the conventional formulation expresses pressure as $P = P(\rho, T, X, Z)$, this representation would require differentiating through the EOS when evaluating the pressure gradient in Eq.~\eqref{eq:temperature_time}, increasing computational cost during PINN training.

To avoid this issue, we invert the EOS relation and train an auxiliary neural network to predict density directly from pressure, temperature, and composition:
\begin{equation}
    \hat{\rho} = \hat{\rho}^{\,\mathrm{net}}(\hat{P}, \hat{T}, X, Z).
\end{equation}
This formulation is physically equivalent but ensures that the auxiliary network appears only in algebraic form, avoiding the need for backpropagation through thermodynamic derivatives. The network is trained on tabulated EOS data constructed from standard sources (OPAL, SCVH, HELM, and PC) \cite{rogers2002updated, saumon1995equation, timmes2000accuracy, potekhin2010thermodynamic}, blended following established procedures \cite{paxton2011modules}. To accurately capture the sharp gradients and multi-scale structure present in EOS tables, the network employs Random Fourier Feature (RFF) \cite{rahimi2007random} embeddings in the input layer together with a compact multilayer perceptron using sinusoidal activations. This design mitigates the spectral bias \cite{jacot2018neural} of standard multilayer perceptrons (MLPs), enabling efficient representation of high-frequency variations without requiring large network capacity. The RFF parameterization is implemented following the approach of \cite{shi2024adaptive}. Once trained, the auxiliary model provides a smooth and differentiable surrogate that replaces traditional interpolation routines. Implementation details regarding table blending and network configuration are provided in the Supplementary Material (Section 4).

\subsubsection{Opacity as an Auxiliary Network}
\label{sec:opacity}
The opacity $\kappa$ governs the efficiency of radiative energy transport and enters the temperature equation through the radiative gradient. The total opacity combines radiative and conductive contributions via the harmonic sum
\begin{equation}
    \frac{1}{\kappa} = \frac{1}{\kappa_{\mathrm{rad}}} + \frac{1}{\kappa_{\mathrm{cond}}}.
\end{equation}

Following the same strategy as for the EOS, we train a second auxiliary neural network to produce a surrogate differentiable model that reproduces the discrete tabulated opacity data as a function of the thermodynamic state:
\begin{equation}
    \kappa = \kappa^{\mathrm{net}}(\hat{P}, \hat{T}, X, Z).
\end{equation}

Parameterizing $\kappa$ in terms of $(\hat{P}, \hat{T})$ ensures consistency with the EOS network and avoids additional coupling during training. The architecture mirrors that of the EOS surrogate, including Fourier feature embeddings and sinusoidal activations, and is trained on opacity tables constructed from standard radiative and conductive sources. Details of the tabulated data construction of $\kappa$ and training procedure are provided in the Supplementary Material (Section 2), which is dedicated to the detailed analysis of the energy transport.

\begin{figure*}
    \centering
    \includegraphics[width=0.7\linewidth]{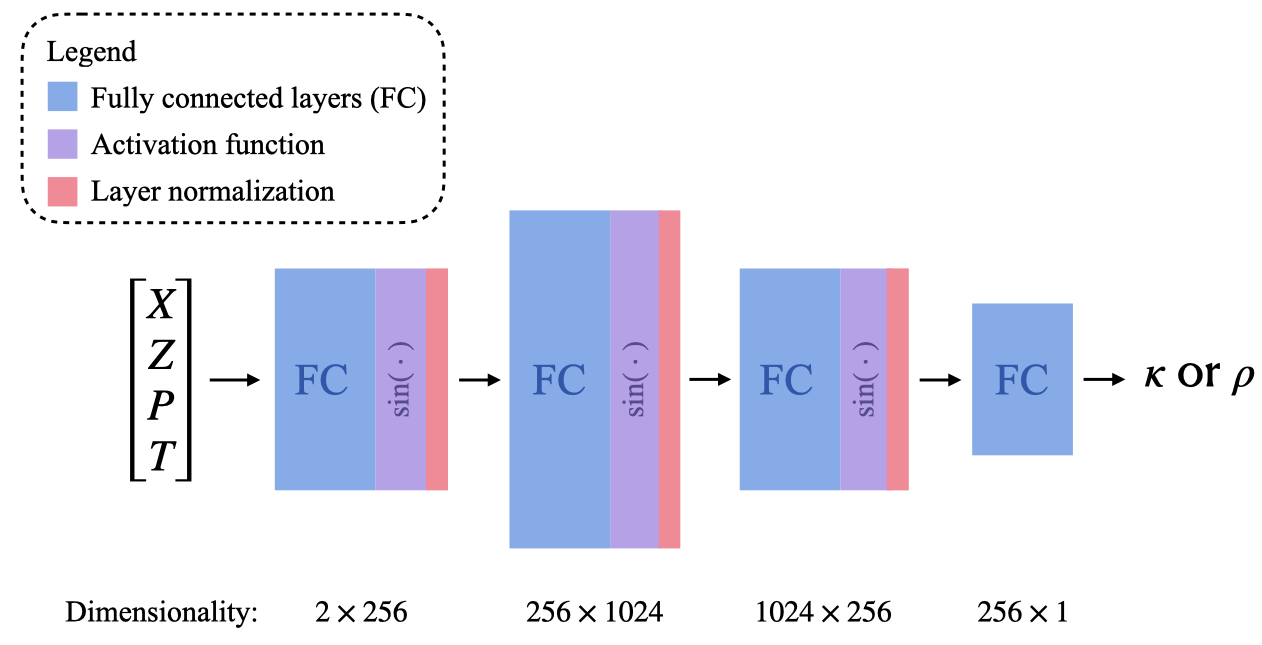}
    \caption{Architecture of the auxiliary network used to learn thermodynamic closures. The same architecture is employed for both the EOS (predicting $\rho$) and opacity ($\kappa$), enabling smooth and differentiable approximations of tabulated microphysics.}
    \label{fig:auxillary_architecture}
\end{figure*}

\subsubsection{Energy Generation}
\label{sec:energy_generation}
The local energy generation rate entering Eq.~\eqref{eq:luminosity_time} is given by
\begin{equation}
    \epsilon = \epsilon_{\mathrm{nuc}} - \epsilon_{\nu} + \epsilon_{\mathrm{grav}},
\end{equation}
where $\epsilon_{\mathrm{nuc}}$ represents nuclear energy release, $\epsilon_{\nu}$ accounts for thermal losses due to neutrinos, and $\epsilon_{\mathrm{grav}}$ captures energy exchange due to gravitational contraction or expansion. In the general time-dependent formulation, the gravitational contribution is directly related to entropy evolution,
\begin{equation}
    \epsilon_{\mathrm{grav}} = -\,T \frac{\partial S}{\partial t},
\end{equation}
which reflects the conversion between thermal and gravitational energy. Under hydrostatic and thermal equilibrium, the entropy is time-independent and $\epsilon_{\mathrm{grav}} \approx 0$.

The nuclear energy generation rate $\epsilon_{\mathrm{nuc}}(\rho, T, X_i)$ exhibits an extreme sensitivity to temperature and spans many orders of magnitude across the stellar interior, reflecting the underlying nuclear reaction networks and Coulomb barrier penetration effects. In addition, neutrino losses $\epsilon_\nu(\rho, T, X_i)$ introduce further nonlinear and regime-dependent behavior. As a result, the total energy generation rate $\epsilon$ is a highly stiff function of the local thermodynamic state.

Neural-network surrogates for stellar microphysics, including nuclear energy generation and equation-of-state quantities, have been developed in recent work, particularly in the context of stellar modeling and emulation \cite{bellinger2016asteroseismic, verma2021machine}. However, incorporating such surrogates within a physics-informed neural network framework introduces additional challenges. In particular, the strong stiffness and sharp local variations of $\epsilon$ can adversely affect training stability and significantly increase computational cost when coupled to the global PDE constraints.

For these reasons, we evaluate the energy generation rate using the microphysical routines from \texttt{MESA}, which compute nuclear reaction rates and neutrino losses based on established physics and finite-difference schemes (see Section 5 of the Supplementary Material for more details). This hybrid approach preserves physical fidelity for the most complex microphysical processes while allowing the neural network to focus on learning the global structure of the stellar solution.

\section{Physics-Informed Neural Network Formulation}

\subsection{Network architecture and physics-informed loss}
We construct a PINN model that directly approximates the solution of the steady-state stellar structure equations. The model is based on a SIREN (Sinusoidal Representation Network) architecture \cite{sitzmann2020implicit}, implemented as a fully connected multi-layer perceptron with sinusoidal activation functions.

The network takes the normalized enclosed mass coordinate $\hat{M}_r$ as input and outputs continuous approximations of the stellar variables,
\[
\left\{
\hat{P}^{\mathrm{net}}(\hat{M}_r;W),\;
\hat{r}^{\mathrm{net}}(\hat{M}_r;W),\;
\hat{T}^{\mathrm{net}}(\hat{M}_r;W),\;
\hat{L}_r^{\mathrm{net}}(\hat{M}_r;W)
\right\},
\]
where $W$ denotes the trainable parameters of the network.

This compact architecture is sufficient to represent the smooth yet highly nonlinear stellar profiles while maintaining a computationally efficient evaluation of derivatives. The use of sinusoidal activations enables accurate representation of high-frequency features and, very importantly, ensures stable gradient backpropagation. This is particularly advantageous in physics-informed settings where one has to calculate the derivatives of the output with respect to the input to evaluate the PDE residual.

The network is trained in a fully self-supervised manner by minimizing the residuals of the governing equations. These residuals are evaluated at a set of collocation points $\{\hat{M}_r^{(i)}\}_{i=1}^{N_c}$ sampled within the stellar interior. The residuals corresponding to the stellar structure equations are defined as
\begin{align}
\mathcal{R}_P &= 
\frac{\partial \hat{P}^{\mathrm{net}}}{\partial \hat{M}_r}
+ \frac{\beta_a \hat{M}_r}{\left(\hat{r}^{\mathrm{net}}\right)^4}, \\[6pt]
\mathcal{R}_r &= 
\frac{\partial \hat{r}^{\mathrm{net}}}{\partial \hat{M}_r}
- \frac{\beta_b}{\left(\hat{r}^{\mathrm{net}}\right)^2 \hat{\rho}}, \\[6pt]
\mathcal{R}_T &= 
\frac{\partial \hat{T}^{\mathrm{net}}}{\partial \hat{M}_r}
+ \frac{\beta_c \hat{L}_r^{\mathrm{net}}}{\left(\hat{r}^{\mathrm{net}}\right)^4}
\, \nabla(\hat{M}_r), \\[6pt]
\mathcal{R}_L &= 
\frac{\partial \hat{L}_r^{\mathrm{net}}}{\partial \hat{M}_r}
- \beta_d \, \epsilon,
\end{align}
where $\hat{\rho}$, $\kappa$, and $\epsilon$ are obtained from the microphysical closures described in Sec.~\ref{sec:microphysics}. The temperature gradient $\nabla$ incorporates both radiative and convective transport through the piecewise formulation introduced in Eq.~\eqref{eq:nabla_piecewise}.

The physics-informed loss is defined as the empirical $L^2$ norm of these residuals,
\begin{equation}
\mathcal{L}_{\mathrm{PDE}}(W)
= \frac{1}{N_c} \sum_{i=1}^{N_c}
\sum_{k \in \{P,r,T,L\}}
\alpha_k \left| \mathcal{R}_k(\hat{M}_r^{(i)};W) \right|^2,
\end{equation}
where $\alpha_k$ are weighting coefficients balancing the contribution of each equation.

The derivatives of the network outputs with respect to $\hat{M}_r$ are computed via automatic differentiation. To improve numerical stability across the wide dynamic range of stellar variables, Layer Normalization is applied before each hidden layer.

\subsection{Hard imposition of boundary conditions}
The stellar structure equations define a boundary-value problem with conditions specified at both the stellar center and surface. In the absence of data-driven supervision, minimizing the loss function $\mathcal{L}_{\mathrm{PDE}}$ alone does not uniquely determine a solution, and boundary conditions must be explicitly enforced. We therefore must impose central regularity conditions and surface boundary conditions obtained from an atmospheric model (the specific boundary values are detailed in Section 3 of the Supplementary Material).

A common approach in PINNs is to impose boundary conditions through soft constraints by augmenting the loss function. However, in fully self-supervised settings, this strategy often leads to slow convergence and sensitivity to the relative weighting of loss terms.

Instead, we adopt a hard-constraint formulation in which the boundary conditions are satisfied exactly by construction. The raw network outputs are transformed using analytic envelope functions that interpolate between the center and surface values,
\begin{equation}
g_u(\hat{M}_r;W) = c_1(m)\, f_u(\hat{M}_r;W)
+ c_2(m)\, u_s + c_3(m)\, u_c,
\end{equation}
where $m = \hat{M}_r / \hat{M}_{\mathrm{total}} \in [0,1]$, $f_u$ denotes the unconstrained network output, and $u_c$, $u_s$ are the prescribed boundary values.

\begin{figure}[htbp]
\centering
\includegraphics[width=0.6\linewidth]{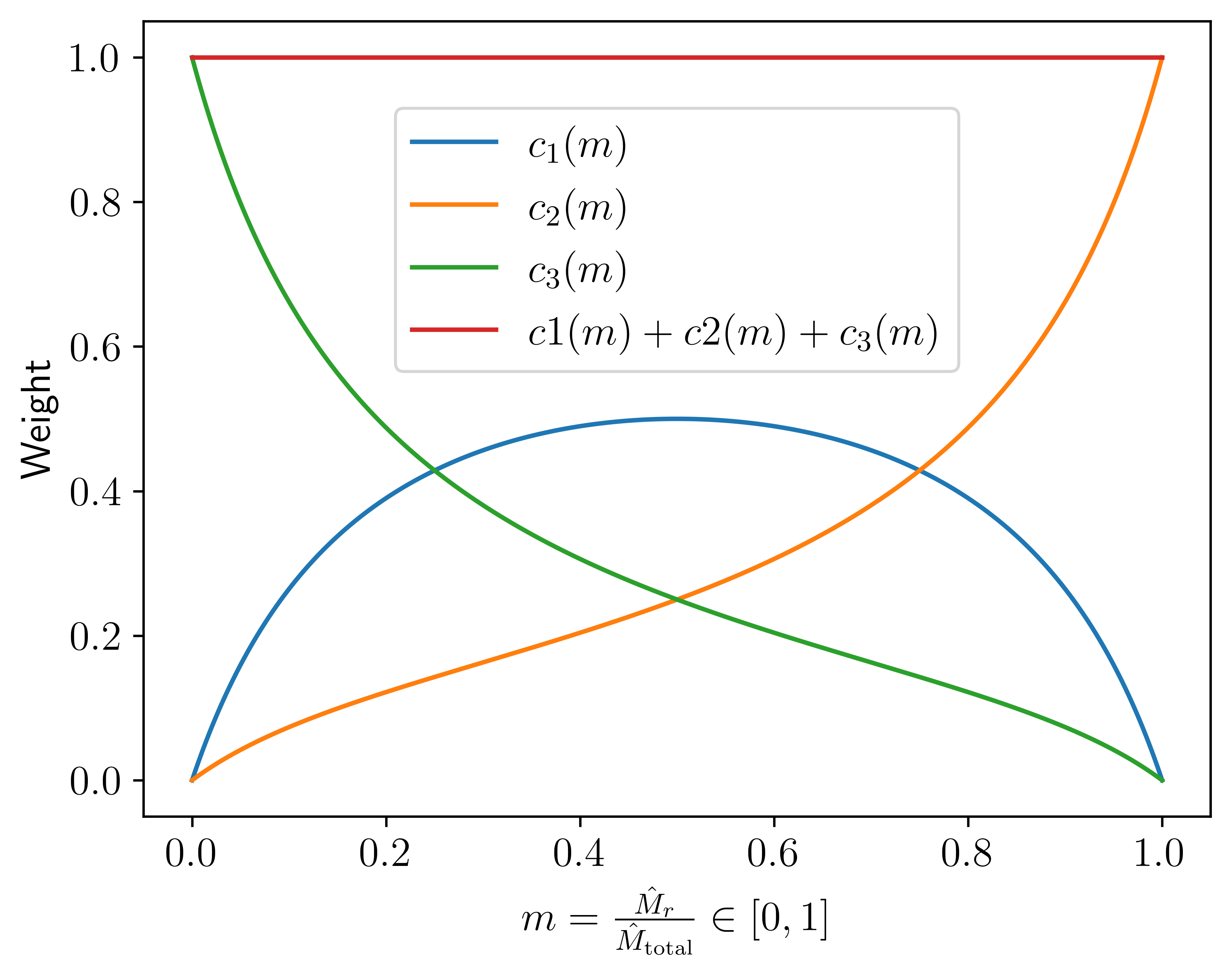}
\caption{\textbf{Stellar profile predictions for relatively low- and high-mass stars.} Each subfigure compares the ground-truth solution obtained from the classical \texttt{MESA} solver (blue) with the PINN prediction (red) for the normalized luminosity, pressure, radius, density, and temperature as functions of enclosed mass.}
\label{fig:hard_constraints}
\end{figure}

The weighting functions are defined as
\begin{align}
c_1(m) &= 1 - \frac{1}{4(m - m^2) + 1}, \nonumber \\
c_2(m) &= \frac{m}{4(m - m^2) + 1}, \nonumber \\
c_3(m) &= \frac{1 - m}{4(m - m^2) + 1},
\end{align}
which ensure that $g_u(0) = u_c$ and $g_u(1) = u_s$ exactly, while remaining smooth and differentiable across the domain. These graphs are shown in Figure \ref{fig:hard_constraints}.

This construction restricts the optimization to the physically admissible solution space, eliminating the need for additional boundary-loss terms and significantly improving convergence stability.

\section{Training procedure and results}
To ensure robust convergence and generalization, we performed an extensive hyperparameter grid search, selecting configurations based on both validation performance and physical consistency. Guided by this process, the final training setup consists of 10,000 iterations with a batch size of 256. The physics-informed loss $\mathcal{L}_p$ is also used as an effective criterion for early stopping.

Optimization is carried out using the Adam optimizer \cite{kingma2014adam} with standard momentum parameters $(\beta_1=0.9, \beta_2=0.999)$ and a weight decay of $10^{-6}$. The learning rate follows a cosine annealing schedule \cite{loshchilov2016sgdr}, initialized at $5\times10^{-4}$ and decaying to $5\times10^{-7}$. For interpolation experiments, the physics-informed loss is scaled by a factor $\lambda=5\times10^{-2}$.

\begin{figure}[htbp]
\centering
\includegraphics[width=0.95\linewidth]{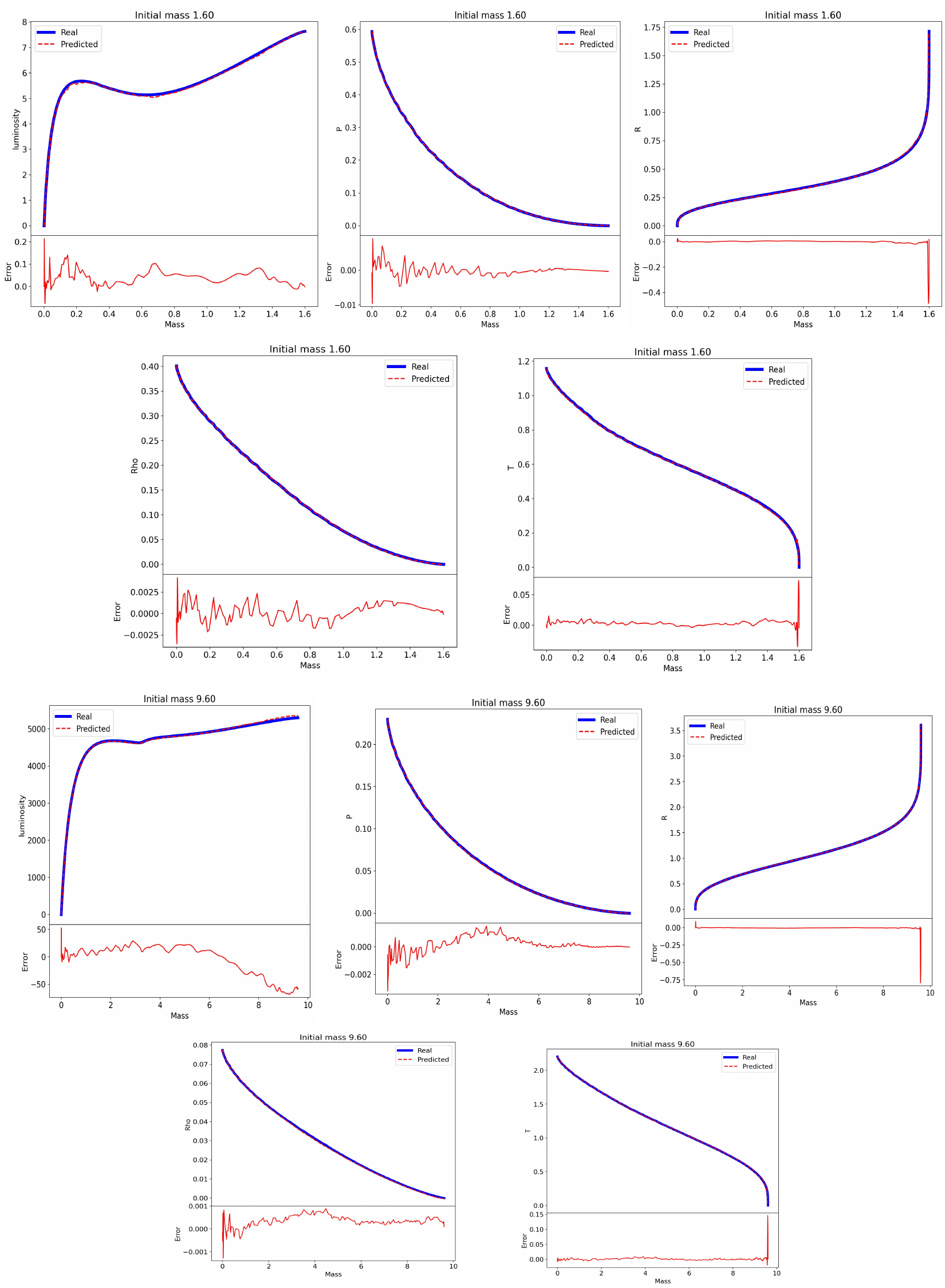}
\caption{\textbf{Stellar profile predictions for relatively low- and high-mass stars.} Each subfigure compares the ground-truth solution obtained from the classical \texttt{MESA} solver (blue) with the PINN prediction (red) for the normalized luminosity, pressure, radius, density, and temperature as functions of enclosed mass.}
\label{fig:profiles}
\end{figure}

To accelerate training and alleviate the computational burden associated with automatic differentiation, we adopt the Stochastic Projection Physics-Informed Neural Network (SP-PINN) framework \cite{navaneeth2023stochastic}. Instead of explicitly computing derivatives through backpropagation, this method approximates the PDE gradients via a Monte Carlo evaluation of nearby collocation points. In practice, we find that sampling a single neighboring point provides sufficient accuracy. 

To further improve efficiency, training batches are constructed such that each collocation point and its corresponding neighbor are included within the same batch. This enables the computation of all required quantities with a single forward pass, eliminating redundant evaluations and significantly reducing both memory usage and runtime.

Collocation points are sampled using the residual-based attention (RBA) strategy \cite{anagnostopoulos2024residual, ramirez2025residual} as a type of active learning. This approach assigns importance weights to points based on the history of their PDE residuals, allowing the model to focus on regions where the governing equations or boundary conditions are not yet well satisfied. Importantly, this mechanism operates without requiring additional gradient computations, making it computationally efficient.

Figure~\ref{fig:profiles} shows the predicted stellar profiles for representative low- and high-mass stars with total masses $M_T = 1.6\,M_\odot$ and $M_T = 9.6\,M_\odot$. The PINN accurately reproduces the reference solutions from \texttt{MESA} across all physical quantities, demonstrating its ability to capture both smooth trends and sharp transitions within the stellar interior.

We evaluate the model across a range of stellar masses from $0.4$ to $9.9\,M_\odot$, using 96 uniformly spaced samples. All models are taken at a stage where approximately $99\%$ of hydrogen has been burned, ensuring a stable equilibrium configuration as provided by \texttt{MESA}. Within this range, the model achieves an average Mean Relative Absolute Error (MRAE) of $3.06\%$ and an average $R^2$ score of $99.98\%$.

The MRAE between a ground-truth quantity $Q_i$ and a prediction $\hat{Q}_i$ is defined as
\begin{equation}
\mathrm{MRAE}(\mathbf{Q},\hat{\mathbf{Q}}) = \frac{1}{N}\sum_{i=1}^N \frac{|Q_i - \hat{Q}_i|}{\sigma(Q_i)},
\label{eq:mrae}
\end{equation}
which provides a normalized and interpretable measure of error across variables with different scales. Unlike metrics such as MSE, RMSE, or MAE, the MRAE captures \textit{relative} discrepancies and therefore better reflects the physical accuracy of the solution.

\begin{figure}[htbp]
\centering
\includegraphics[width=\linewidth]{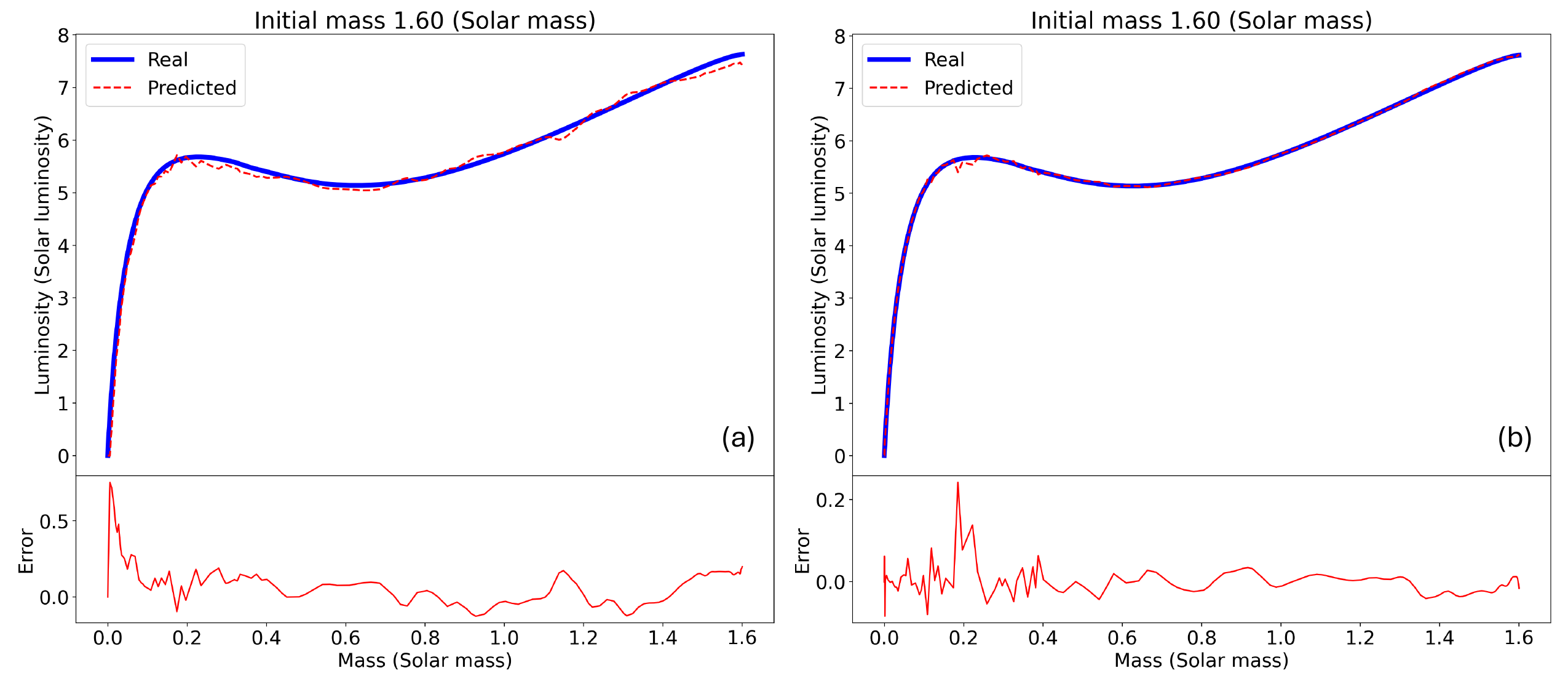}
\caption{\textbf{Fully supervised model with limited data.} Performance of the same architecture but now (a) trained purely on \texttt{MESA} data without enforcing the governing equations, and (b) trained with on \texttt{MESA} data and the governing equations. These models behave as intelligent interpolator rather than solvers.}
\label{fig:supervised}
\end{figure}

From a broad perspective, a PINN combines a physics-based loss with an optional data-driven term,
\begin{equation}
\mathcal{L}(W) = \alpha_{\mathrm{PDE}} \mathcal{L}{\mathrm{PDE}}(W) + \alpha{\mathrm{DD}} \mathcal{L}{\mathrm{DD}}(W),
\end{equation}
while hard constraints are imposed through the architecture. When $\alpha{\mathrm{DD}} > 0$, the model effectively operates in a supervised regime, leveraging known input-output pairs and behaving as a physics-regularized interpolator. To highlight the importance of incorporating physical constraints, we compare our model performance with two additional models (using the same architecture and number of epochs): one trained purely on \texttt{MESA} data without enforcing the governing equations (Figure~\ref{fig:supervised}a), and another trained on \texttt{MESA} data together with the governing equations (Figure~\ref{fig:supervised}a). It should be emphasized that these two comparative models behave as intelligent interpolators rather than solvers, since the solution is known beforehand at specific points.

We carried out this supervised training using 10\% of the original \texttt{MESA} data (the full dataset for the stars under analysis contains around 4,000 track points), with the remaining data used for testing; this fraction was sufficient to achieve proper convergence. In the absence of the PDE loss ($\alpha_{\mathrm{PDE}}=0$), as shown in Figure~\ref{fig:supervised}a, the network exhibits significantly larger errors as well as non-physical oscillations. Using qualitative metrics for these supervised interpolator models, the purely data-driven model achieved an average MRAE of 3.85\%, while incorporating the governing equations reduced this to 2.05\%. For comparison, the self-supervised, data-free solver model achieved 3.06\%. 

While the best performance is obtained when combining both data and governing equations, strongly enforcing the equations alone can yield comparable (or in this case even better) performance than using data alone. This improvement can be attributed to the fact that the data are limited to the particular predefined points, whereas the governing equations are enforced through collocation points that can be freely and adaptively selected across the continuous domain through active learning to minimize the error, providing a stronger and more uniform constraint on the solution. This comparison emphasizes that enforcing the governing equations is essential for obtaining smooth and physically consistent solutions.

As mentioned, our formulation sets $\alpha_{\mathrm{DD}} = 0$, resulting in a fully self-supervised model that acts as an independent solver of the stellar structure equations. Under the appropriate boundary conditions imposed through the architecture, the underlying PDE system admits a unique solution, making the problem well-posed. In this setting, the PINN can be interpreted as a mesh-free numerical solver, analogous in spirit to finite-difference or finite-volume methods, but now with the advantage of producing continuous and differentiable solutions across the domain.

\begin{figure}[!h]
\centering
\includegraphics[width=\linewidth]{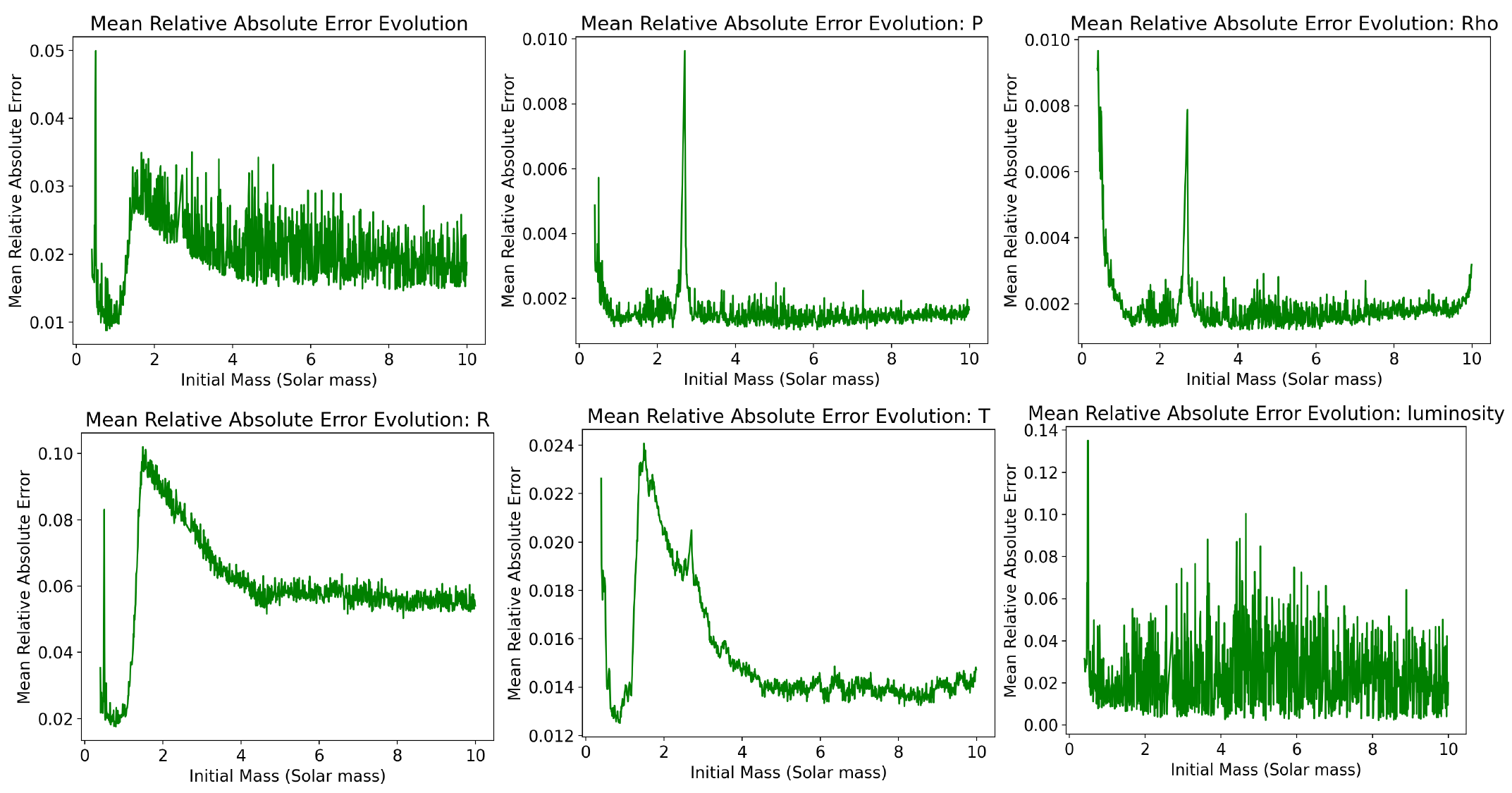}
\caption{\textbf{Performance across stellar masses.} Evaluation of the PINN model for stars with initial masses ranging from $0.4$ to $9.9\,M_\odot$. The plot shows the MRAE for each physical quantity as well as the total error.}
\label{fig:errors}
\end{figure}

The performance across different stellar masses is summarized in Figure~\ref{fig:errors}. The MRAE remains relatively uniform across the studied range, although lower-mass stars exhibit higher variance. This behavior is consistent with their increased sensitivity to stability conditions near the chosen evolutionary stage.

Finally, we investigate extending the model to time-dependent stellar evolution by including time as an additional input. In this setting, the model produces reasonable estimates for global quantities such as the effective temperature $T_{\mathrm{eff}}$ and luminosity $L_{\mathrm{eff}}$, which capture integrated properties of the stellar profile. However, the internal structure predictions become significantly noisier and less accurate.

\begin{figure}[!h]
\centering
\includegraphics[width=0.7\linewidth]{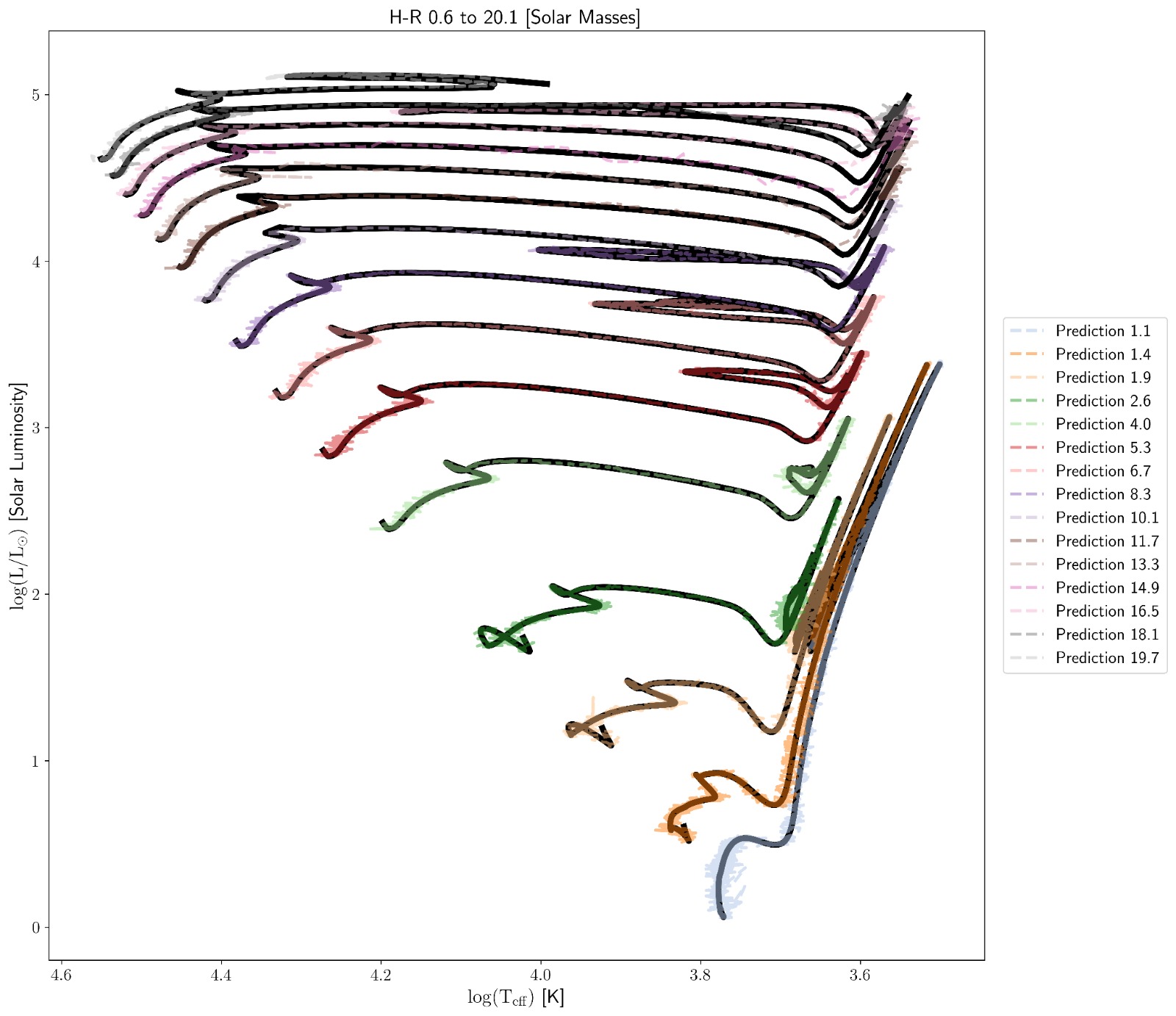}
\caption{\textbf{Hertzsprung–Russell diagram.} Comparison between the proposed PINN model (dashed lines) and \texttt{MESA} simulations (solid lines) for stars with initial masses between $0.6$ and $20\,M_\odot$. The x-axis shows $\log(T_{\mathrm{eff}})$ and the y-axis $\log(L/L_\odot)$. While the global trends are captured, noticeable noise and deviations highlight the limitations of the current model in time-dependent settings.}
\label{fig:HR}
\end{figure}

This behavior is illustrated in Figure~\ref{fig:HR}, which shows the Hertzsprung–Russell diagram for stars with initial masses between $0.6$ and $20\,M_\odot$. While the overall trends are qualitatively captured, high-frequency noise and deviations from the reference solutions are evident. Our results indicate that the present formulation does not readily extend to time-dependent problems, and that more specialized approaches are required.

\section{Discussion and conclusions}
In this work, we present a first demonstration of an self-supervised PINN framework tailored to the stellar structure equations, combining several recent advances in the literature into a unified and physically consistent model. Standard PINNs, when applied directly, struggle to accurately reproduce stellar structure due to the stiffness of the equations, the strong coupling between variables, and the strict boundary conditions required at the stellar center and surface. 

To address these limitations, we introduce a set of complementary modifications. First, the use of physics-based loss terms combined with architectural transformations that enforce boundary conditions as hard constraints ensures that the learned solutions remain physically admissible throughout training. Second, auxiliary neural networks are employed to replace traditional tabulated microphysics (equation of state and opacity) with smooth and differentiable surrogates, enabling fully end-to-end training. The inclusion of Random Fourier Features (RFF) in these auxiliary models is essential to capture the sharp gradients present in the tabulated data, mitigating the spectral bias of standard MLPs.

A key design consideration throughout this work is the trade-off between expressivity and computational efficiency. Both the auxiliary networks and the main PINN were carefully optimized to remain as compact as possible while still capturing the relevant physical behavior. While RFF embeddings proved critical for the auxiliary models, we found that using a SIREN architecture for the main PINN provides a better balance, enabling the representation of high-frequency features with fewer parameters and faster evaluation.

Another important contribution is the integration of the Stochastic Projection PINN (SP-PINN) framework, which enables gradient-free approximation of derivatives required for the PDE residuals. By estimating derivatives from nearby collocation points, this approach significantly reduces the computational overhead associated with automatic differentiation, leading to faster training and lower memory usage. In addition, the use of a residual-based attention strategy for active learning further improves performance by concentrating collocation points in regions where the solution is more complex or less well-resolved.

Overall, the proposed framework produces accurate and smooth stellar profiles in hydrostatic and thermal equilibrium, capturing both radiative and convective transport regimes, as well as realistic energy generation rates computed from established microphysics. The agreement with classical finite-difference solvers such as \texttt{MESA} demonstrates that PINNs can serve as a viable alternative for modeling stellar interiors, while offering additional advantages such as differentiability and mesh-free evaluation.

Despite these promising results, several limitations remain. Most notably, the current formulation is restricted to time-independent (equilibrium) stellar structure. Extending the model to include time evolution is non-trivial. Preliminary experiments show that directly adding time as an additional input leads to reasonable predictions for global quantities such as $T_{\mathrm{eff}}$ and $L_{\mathrm{eff}}$ (used for the HR plot) but fails to accurately reproduce the internal structure, introducing high-frequency noise. This indicates that the present architecture does not readily generalize to time-dependent problems. Future work may therefore explore hybrid approaches, such as combining finite-difference schemes in time with neural representations in space, or developing specialized architectures designed for evolutionary dynamics. Additionally, extending the auxiliary modeling of the energy generation rate $\epsilon$ to include time-dependent effects (with the gravitational term) could further improve consistency for evolving stars. With the addition of time, there should also be a focus on extending the analysis to the mass range $0.4$–$10\,M_\odot$ to more extreme regimes (including very low-mass stars and high-mass stars approaching supernova conditions).

In summary, this work establishes a foundation for data-free, physics-informed neural modeling of stellar interiors, opening the door to scalable and differentiable simulations for large-scale astrophysical applications.

\bibliography{sample}

\section*{Author contributions statement}
M.B., S.L.P, S.G., C.W., V.K. and A.K.K conceived the original idea. M.B. developed the methodology and wrote the manuscript draft. S.L.P. implemented the PINN program and wrote part of the refined manuscript. S.G. run the \texttt{MESA} models and wrote part of the refined manuscript. P.K. wrote the original sections related to the energy generation rate and refined the final manuscript. P.M.S. and U.D. interpreted the results and provided visualizations in the manuscript. A.M. performed a dimensional mathematical analysis of the stellar equations. C.W. and S.C. improved the PINN model and revised the manuscript. V.K. and A.K.K corrected the manuscript and supervised the project. Y.J. restructured the refined manuscript. All the authors attended discussions and provided relevant insights. All authors reviewed and approved the manuscript.

\section*{Additional information}
\noindent \textbf{Acknowledgment:} The authors gratefully acknowledge support from the NSF-Simons AI Institute for the Sky (SkAI), funded by the U.S. National Science Foundation and the Simons Foundation.

This research used the DeltaAI advanced computing and data resource, which is supported by the National Science Foundation (award OAC 2320345) and the State of Illinois. DeltaAI is a joint effort of the University of Illinois Urbana-Champaign and its National Center for Supercomputing Applications.

\end{document}


\maketitle

\section{Governing equations of stellar structure}
\label{sec:stellar-equations}
We consider a spherically symmetric, self-gravitating star whose internal state is described by the density $\rho$, temperature $T$, pressure $P$, and chemical composition $(X_i)$. All thermodynamic and transport quantities are functions of $(\rho, T, X_i)$, and the system is governed by conservation of mass, momentum, and energy, together with a prescription for energy transport.

\subsection{Eulerian formulation}
The stellar structure equations can be written as
\begin{align}
& \frac{1}{\rho} \frac{dP}{dr} + \frac{G M_r}{r^2} + \frac{d^2 r}{dt^2} = 0, 
\label{eq:hydro-r} \\[4pt]
& \frac{dM_r}{dr} = 4\pi r^2 \rho,
\label{eq:mass-r} \\[4pt]
& \frac{dT}{dr} = -\,\frac{G M_r T}{4\pi r^4 P}\,\nabla(r,t),
\label{eq:temp-r} \\[4pt]
& \frac{dL_r}{dr} = 4\pi r^2 \rho \left(\varepsilon_{\rm nuc} - \varepsilon_{\nu} + \varepsilon_{\rm grav}\right),
\label{eq:lum-r}
\end{align}
where $M_r$ is the enclosed mass and $L_r$ is the luminosity crossing a sphere of radius $r$. The source terms represent nuclear energy generation ($\varepsilon_{\rm nuc}$), thermal neutrino losses ($\varepsilon_{\nu}$), and gravitational energy exchange. The latter is related to entropy evolution through
\begin{equation}
\varepsilon_{\rm grav} = -T \frac{dS}{dt}.
\end{equation}

The dimensionless temperature gradient
\begin{equation}
\nabla = \frac{d \ln T}{d \ln P}
\end{equation}
encodes the efficiency of energy transport and determines whether energy is carried predominantly by radiation or convection. Its modeling is described in detail in Section~2.

\subsection{Microphysics and Closure}
Equations~\eqref{eq:hydro-r}–\eqref{eq:lum-r} provide four differential relations for five unknowns $(M_r, P, \rho, T, L_r)$ and must be supplemented by closure relations. In our formulation:
\begin{itemize}
    \item The opacity $\kappa(P,T,X,Z)$, required to evaluate radiative transport, is also modeled via an auxiliary neural network (Section~2.2).
    \item     The equation of state (EOS) provides $\rho = \rho(P,T,X,Z)$ and is represented by an auxiliary neural network (Section~4).
    \item The energy generation rate $\epsilon$ is evaluated using classical finite-difference microphysical routines (Section~5).
\end{itemize}

This hybrid strategy separates stiff microphysics from the learnable components of the model, improving numerical stability while preserving physical accuracy.

\subsection{Lagrangian formulation}
Since the enclosed mass $M_r$ is a monotonically increasing function of radius, it is convenient to adopt $M_r$ as the independent variable. Using the transformation
\begin{equation}
\frac{d}{dr} = 4\pi r^2 \rho \frac{d}{dM_r},
\end{equation}
the system can be rewritten in Lagrangian form as
\begin{align}
\frac{\partial P}{\partial M_r}
&= -\frac{G M_r}{4\pi r^4} - \frac{1}{4\pi r^2}\frac{\partial^2 r}{\partial t^2}, \\
\frac{\partial r}{\partial M_r}
&= \frac{1}{4\pi r^2 \rho}, \\
\frac{\partial T}{\partial M_r}
&= -\frac{3\kappa L_r}{64\pi^2 a c\, T^3 r^4}, \\
\frac{\partial L_r}{\partial M_r}
&= \epsilon - T \frac{\partial S}{\partial t}.
\end{align}

In this work, we restrict attention to quasi-static stellar configurations in hydrostatic and thermal equilibrium, such that
\begin{equation}
\frac{\partial^2 r}{\partial t^2} \approx 0, 
\qquad 
\frac{\partial S}{\partial t} \approx 0,
\end{equation}
reducing the system to a set of coupled ordinary differential equations.

\subsection{Non-dimensional formulation}
To improve numerical conditioning and facilitate training of the physics-informed neural network, we introduce dimensionless variables normalized by solar reference values:
\begin{equation}
\hat{M}_r = \frac{M_r}{M_\odot}, \quad
\hat{r} = \frac{r}{R_\odot}, \quad
\hat{P} = \frac{P}{P_\odot}, \quad
\hat{T} = \frac{T}{T_\odot}, \quad
\hat{L}_r = \frac{L_r}{L_\odot}.
\end{equation}

This scaling follows standard dimensional analysis in stellar structure, where characteristic pressure, temperature, and density scales can be estimated from global stellar properties and hydrostatic balance. Substituting these variables into the equilibrium equations yields, for the case of radiative transfer, the following expression
\begin{subequations}
\label{eq:scaled_time_independent_mass_structure}
\begin{align}
\frac{\partial \hat{P}}{\partial \hat{M}_r}
  &= \frac{\beta_a\hat{M}_r}{\hat{r}^4},\\ 
\frac{\partial \hat{r}}{\partial \hat{M}_r}
  &= \frac{\beta_b}{\hat{r}^2\hat{\rho}}, \\
\frac{\partial \hat{T}}{\partial \hat{M}_r}
  &= \frac{\beta_c\kappa \hat{L}_r}{\hat{T}^3 \hat{r}^4}, \\
\frac{\partial \hat{L}_r}{\partial \hat{M}_r}
  &= \beta_d\epsilon,
\end{align}
\end{subequations}
where the dimensionless coefficients are
\begin{align}
\beta_a &= -\frac{GM^2_\odot}{4\pi P_\odot R_\odot^4}, \quad
\beta_b =  \frac{M_\odot}{4\pi R^3_\odot\rho_\odot}, \nonumber\\
\beta_c &=  -\frac{3M_\odot L_\odot}{64\pi^2acT_\odot^4R^4_\odot}, \quad
\beta_d = \frac{M_\odot}{L_\odot}.
\end{align}

Under solar normalization, these coefficients reduce the dynamic range of the problem and improves optimization in the PINN framework. Observe that this is the dimensionless system corresponding to the governing equations enforced by the physics-informed network described in the main text.

\section{Energy transport modeling}
\label{sec:transport}
The luminosity enclosed within radius $r$ determines the net outward photon energy flux. In spherical coordinates,
\begin{equation}
F_r = \frac{L_r}{4\pi r^2},
\label{eq:Fr_def}
\end{equation}
represents the flux transported through the stellar interior by radiative diffusion, convection, or a combination of both. The temperature gradient $\nabla(r,t) = d\ln T / d\ln P$ from the stellar structure equations determines which mechanism dominates.

\subsection{Radiative transport}
\label{subsec:radiative_opacity}
Photons propagate outward in a relatively transparent medium through radiative diffusion, following a random walk due to repeated absorption and re-emission. The corresponding temperature gradient is
\begin{equation}
\nabla_{\rm rad}
= \frac{3\,\kappa\,P\,L_r}{16\pi a c G M_r T^4},
\label{eq:nabla-rad}
\end{equation}
where $a$ is the radiation constant and $c$ is the speed of light. The quantities $P(r)$, $L_r(r)$, $M_r(r)$, and $T(r)$ are provided by the stellar structure equations, while the opacity $\kappa(\rho,T,X,Z)$ encapsulates the microphysical properties of the stellar material and constitutes the key additional input required to determine radiative energy transport, as discussed below. For completeness, the derivation of Eq.~\ref{eq:nabla-rad} from Eq.~\ref{eq:Fr_def} is presented below.

We now derive, for the completeness of the manuscript, the standard expression for the radiative temperature gradient, $\nabla_{\rm rad}$ \cite{pols2011stellar}. In spherical symmetry, the luminosity enclosed within radius $r$, denoted $L_r$, is defined as the total energy per unit time crossing a sphere of radius $r$, as given in Eq.~\ref{eq:Fr_def}. Deep in stellar interiors, the medium is typically optically thick, so radiative transfer can be treated in the \textit{diffusion} approximation. In this limit, the radiative energy density is given by
\[
u_{\rm rad} = aT^4,
\]
where $a$ is the radiation constant.

From radiative transport theory, the radiative energy flux may be written
\begin{equation}
\mathbf{F}_{\rm rad} = -D\,\nabla u_{\rm rad} = -\frac{4acT^3}{3\kappa\rho}\,\nabla T, 
\label{eq:diffusion_general}
\end{equation}
where $D = c/(3\kappa\rho)$, and the factor $1/3$ arises from averaging over photon directions in an approximately isotropic radiation field. Under spherical symmetry, only the radial component is nonzero, so
\begin{equation}
\frac{L_r}{4\pi r^2} = F_{\rm rad}
= -\frac{4acT^3}{3\kappa\rho}\,\frac{dT}{dr}.
\label{eq:Frad_diffusion}
\end{equation}

We now express $dT/dr$ in terms of $\nabla$ under hydrostatic equilibrium. Using $\nabla = d\ln T/d\ln P$ and applying the chain rule,
\begin{equation}
\frac{dT}{dr}
= T\,\nabla_{\rm rad}\,\frac{1}{P}\,\frac{dP}{dr}.
\label{eq:dTdr_chain}
\end{equation}

The radial derivative of the pressure at hydrostatic equilibrium is given by 
\begin{equation}
\frac{dP}{dr} = -\frac{G M_r \rho}{r^2},
\end{equation}
which, when substituted into Eq.~\ref{eq:dTdr_chain}, yields
\begin{equation}
\frac{dT}{dr}
= -\,\frac{G M_r \rho}{r^2}\,\frac{T}{P}\,\nabla_{\rm rad}.
\label{eq:dTdr_hse}
\end{equation}
Substituting this expression into Eq.~\ref{eq:Frad_diffusion} leads to
\begin{equation}
\nabla_{\rm rad} =
\frac{3\kappa P L_r}{16\pi acG M_r T^4}.
\end{equation}

Whether radiation alone can transport the energy flux depends on the balance between the radiative temperature gradient and the adiabatic gradient $\nabla_{\rm ad}(r,t)$, which varies across stellar masses. A region is convectively stable when $\nabla_{\rm rad}(r,t) \le \nabla_{\rm ad}(r,t)$, in which case $\nabla(r,t) = \nabla_{\rm rad}(r,t)$. Physically, radiative diffusion is most efficient in regions where the gas is highly ionized and the opacity correspondingly low, allowing photons to transport energy outward with relatively small temperature gradients.

\subsection{Opacity auxiliary network}
Following the same strategy previously used to learn the density $\rho$ from $(P,T,X,Z)$ with an auxiliary network (trained on tabulated EOS data), we introduce an auxiliary network to model the opacity $\kappa$ as a function of the same four-dimensional input. Before describing the network architecture and training procedure, we briefly outline how the opacity data are constructed.

Radiative opacities are taken from the OPAL \cite{rogers2002updated} tables at high temperatures ($3.75 \le \log T \le 8.7$). Lower temperatures ($2.7 \le \log T \le 4.5$) require a different table \cite{ferguson2005low}, and higher temperatures ($\log T \ge 8.7$) are treated using an analytical formula \cite{buchler1976compton} that accounts for dominant Compton scattering. A smooth overlap of the different components of $\kappa_{\rm rad}$ is performed across the relevant ranges of $\log \rho$ and $\log T$.

Electron conduction opacities are taken from \cite{cassisi2007updated} over the range $-6 \le \log \rho \le 9.75$ and $3 \le \log T \le 9$. Outside these ranges, the tables are extended using \cite{hubbard1969thermal} for non-degenerate electrons and \cite{yakovlev1980thermal} for degenerate electrons.

The total opacity is computed as the harmonic sum
\[
\frac{1}{\kappa} = \frac{1}{\kappa_{\mathrm{rad}}} + \frac{1}{\kappa_{\mathrm{cond}}},
\]
and tabulated as a function of $(\rho, T, X, Z)$. Using the EOS to relate density and pressure, we recast the input domain as $(P, T, X, Z)$, avoiding explicit coupling between auxiliary networks during PINN training. The auxiliary neural network is then trained to learn a smooth surrogate $\kappa(P,T,X,Z)$, following the same architecture and training strategy as the EOS auxiliary network, later described in Sec.~\ref{sec:eos}.

\subsection{The adiabatic gradient}
\label{subsec:adiabatic}
The adiabatic temperature gradient measures how the temperature of a fluid element changes with pressure when displaced without heat exchange, and is defined as
\begin{equation}
\nabla_{\rm ad} = \left(\frac{d\ln T}{d\ln P}\right)_{S}.
\end{equation}
Using thermodynamic identities, it can be written as
\begin{equation}
\nabla_{\rm ad} = \frac{P\,\delta}{\rho\,T\,c_P},
\qquad
\delta = - \left( \frac{\partial \ln\rho}{\partial \ln T} \right)_{P}.
\end{equation}
An equivalent form is
\begin{equation}
\nabla_{\rm ad} = \frac{\Gamma_3 - 1}{\Gamma_1}.
\end{equation}
In practice, \texttt{MESA} evaluates these quantities consistently from EOS tables.

\subsection{Convective transport}
\label{subsec:convective}
In regions where $\nabla_{\rm rad} > \nabla_{\rm ad}$, the medium becomes convectively unstable. The temperature gradient then satisfies $\nabla_{\rm ad} \le \nabla \le \nabla_{\rm rad}$, and we denote $\nabla = \nabla_{\rm conv}$.

Convective transport in \texttt{MESA} \cite{paxton2011modules} is modeled using mixing-length theory (MLT). Convective elements travel a characteristic distance
\begin{equation}
\Lambda(r) = \alpha_{\rm MLT} H_P(r),
\end{equation}
where $H_P = P/(\rho g)$ is the pressure scale height.

The superadiabaticity $\nabla - \nabla_{\rm ad}$ characterizes the buoyancy of a convective plasma element. MESA solves MLT equations to obtain $\nabla_{\rm MLT}(r)$ and the mixing coefficient $D_{\rm conv}(r)$. Convective overshooting is modeled as
\begin{equation}
D_{\rm ov}(z)
=
D_{\rm conv}(r_0)\,
\exp\!\left(-\frac{2z}{f H_{P}(r_0)}\right),
\end{equation}
and affects chemical mixing but not directly the four stellar structure equations. The basic condition for convective energy transport
\begin{equation}
\nabla_{\rm rad} > \nabla_{\rm ad},
\end{equation}
is known as the Schwarzschild criterion. The more refined Ledoux criterion implemented in \texttt{MESA} modifies this inequality to
\begin{equation}
\nabla_{\rm rad} > \nabla_L,
\end{equation}
where $\nabla_L = \nabla_{\rm ad} + B$, and $B$ accounts for the impeding effect of compositional gradients on the motion of convective elements and their ability to transport energy efficiently.

The actual total temperature gradient that accounts for radiative diffusion and convection is approximated as
\begin{equation}
\nabla = (1-\zeta)\nabla_{\rm rad} + \nabla_{\rm ad},
\end{equation}
(or $\nabla_L$ in place of $\nabla_{\rm ad}$ under the Ledoux criterion) where $\zeta$ measures convective efficiency. Convection dominates in high-opacity or high-flux regions, while radiation dominates elsewhere.

\section{Boundary conditions}
\label{sec:bc}
The interior equations must be supplemented by boundary conditions at the center and at the stellar surface.

\paragraph{Central boundary conditions.}
At the center ($r=0$), regularity requires
\begin{equation}
M_r = 0,\qquad L_r = 0.
\end{equation}
The central temperature $T_c$ and density $\rho_c$ are free parameters that may be determined as part of the global solution \cite{pols2011stellar}.

\paragraph{Photospheric/outer boundary conditions.}
While the surface of a star is ill-defined, in practice, a choice must be made, and that is often the photosphere. The photosphere is defined as the point in the stellar atmosphere where the optical depth, $\tau$, reaches a value of 2/3. This is roughly where the stellar plasma becomes opaque to photons, representing the lowest visible layer in the star's atmosphere and effectively its visible ``surface''. In general, a number of reasonable assumptions can be made to set the surface boundary conditions. \texttt{MESA} uses stellar atmosphere models to set the surface boundary conditions \cite{paxton2011modules}. \texttt{MESA} uses the mass, radius, and luminosity to calculate the surface temperature and pressure ($T_{\rm surf}$ and $P_{\rm surf}$) for the model.

Depending on the star, a particular atmosphere model may be more or less appropriate to use and several options are available. They may be either analytic (such as the Eddington T-$\tau$ relation) or numerical. In the former case, an integration of analytical relations is performed over a range of optical depths to obtain $P_{\rm surf}$ and $T_{\rm surf}$. In the latter case, tables are constructed by the PHOENIX \cite{hauschildt1999nextgen, hauschildt1999nextgen2}, by the Castelli and Kurucz model \cite{castelli2003modelling}, or by the COND \cite{allard2001limiting}. These tables are used for different regions and types of stars as detailed in \texttt{MESA} \cite{paxton2011modules}, resulting on various codes used to numerically solve non-local thermodynamic radiative transfer equations under various conditions.

Finally, there is a relatively simple approach taken by \texttt{MESA} to set the boundary conditions, which derives from a constant opacity solution to radiative diffusion:

\begin{equation}
P_{\rm surf} = \frac{\tau_{\rm surf} g}{\kappa_{\rm surf}}\left[1 + 1.6\times10^{-4}\left(\frac{L/L_{\odot}}{M/M_{\odot}}\right)\right]
\end{equation}
where $\kappa_{\rm surf}$ is the constant opacity (that may either be given or else calculated iteratively. Here $\tau_{surf}$ is the surface optical depth which must be specified (often taken as $\tau_{\rm surf} = 2/3$ as discussed above. In this way, $P_{\rm surf}$ may be calculated. To determine $T_{\rm surf}$ in this case, the Eddington $T(\tau)$ relation is used, given $\tau_{\rm surf}$. I.e.,

\begin{equation}
T(\tau) = \frac{3}{4} T^4_\text{eff}\left(\tau + 2/3\right)
\end{equation}
In this relation, $T_\text{eff}$ is the effective temperature, representing the temperature that the star would have if it were a perfect black body. It is given by the Stefan-Boltzmann law:
\begin{equation}
T^4_\text{eff} = \frac{L}{4\pi R^2 \sigma}
\end{equation}
where $\sigma$ is the Stefan-Boltzmann constant.

Together, these central and surface boundary conditions, the EOS, and the microphysics described above define the boundary–value problem that we solve for the stellar structure variables $\{P,\rho,T,L_r,M_r\}(r,t)$.

\section{Equation of State modeling}
\label{sec:eos}
\subsection{Generation of tabulated EOS data}
The equation-of-state (EOS) data are provided in the form of a multidimensional table parameterized by four independent variables $(\log T,\; \log Q,\; X,\; Z)$. Here $(X, Z)$ specify the composition and $\log Q \equiv \log \rho - 2 \log T + 12$ is a reparameterization of the density designed to reduce the dynamic range of the tabulated data. For each input 4-tuple, the EOS returns the total pressure and its constituent components ($P = P_{\rm gas} + P_{\rm rad}$ with $P_{\rm rad} = a T^4/3$), the entropy $S$, the specific heats $c_V$ and $c_P$, the adiabatic exponents $\Gamma_1$ and $\Gamma_3 - 1$, and the standard logarithmic derivatives
\begin{equation}
\chi_T = \left( \frac{\partial \ln P}{\partial \ln T} \right)_{\rho},
\qquad
\chi_\rho = \left( \frac{\partial \ln P}{\partial \ln \rho} \right)_T.
\end{equation}
Because the EOS tables are constructed on a discrete grid, bicubic spline interpolation in $(\log T,\log Q)$ is used to evaluate intermediate values, combined with quadratic interpolation in the composition variables $(X,Z)$. 

In practice, \texttt{MESA} \cite{paxton2011modules} constructs the EOS by combining several well-established reference models, each covering a different region of the input domain. The usual domain involves a metallicity range of $0 < Z < 0.04$ and the rectangular $(\log T,\log Q)$ region defined by $2.1 \le \log T \le 8.2$ and $-10 \le \log Q \le 5.69$. In this regime, the thermodynamic quantities are taken from the OPAL \cite{rogers2002updated} and SCVH \cite{saumon1995equation} tables, both blended smoothly across their overlap.

Outside this usual region, \texttt{MESA} switches to analytic EOS formulations that are evaluated directly on the fly. For models with higher metallicity ($Z > 0.04$), all thermodynamic quantities are computed across the entire $(\log T,\log Q)$ plane from the HELM EOS \cite{timmes2000accuracy} (which includes electron-positron pair production) and the PC EOS \cite{potekhin2010thermodynamic} (appropriate when the plasma approaches crystallization). For a metallicity still in the range $0 < Z < 0.04$ but with points lying outside the main rectangular $(\log T,\log Q)$ domain, the model uses HELM at relatively high temperatures and PC at comparatively lower temperatures but high densities.

\subsection{Auxiliary network for the EOS data}
The PINN formulation relies on thermodynamic quantities provided by the EOS, which are available as discrete, precomputed tabulated data. Rather than directly using the \texttt{MESA} EOS module and its classical interpolation routines during training, we introduce an auxiliary neural network that learns a smooth, continuous surrogate of the EOS from the raw tabulated data described in the previous subsection. This auxiliary model is trained independently beforehand and subsequently embedded into the main PINN to provide thermodynamic closures during the solution of the stellar structure equations.

A natural choice would be to train the auxiliary network to predict the pressure $P$ as a function of $(\rho, T, X, Z)$. However, in the PINN formulation of Eq.~\ref{eq:scaled_time_independent_mass_structure}, derivatives of $P$ with respect to the independent coordinate are required. Back-propagating these derivatives through both the PINN and the auxiliary EOS network would substantially increase computational cost and memory usage. To avoid this problem, we instead invert the EOS relationship and train the auxiliary network to predict the density $\rho$ from $(P, T, X, Z)$. This reformulation is fully equivalent from a physical standpoint, while ensuring that the auxiliary network is only queried for $\rho$, whose derivatives are not required in the PDE system.

The auxiliary network is designed to accurately reproduce the sharp gradients and high-frequency structure present in the EOS tables. To this end, the input variables are first mapped through a Fourier feature embedding, which enriches the input representation with sinusoidal basis functions at various frequency scales. This mapping mitigates the spectral bias of standard MLPs and enables the network to resolve high resolution variations in the EOS without requiring a very deep network with a large parameter space.

The network architecture follows a symmetric bottleneck design: an input layer is mapped to 256 units, followed by hidden layers of 1024 and 256 units, and a final linear layer producing a single scalar output corresponding to $\rho$. Sinusoidal activation functions are used throughout, and Layer Normalization is applied before each fully connected layer to stabilize training across the wide dynamic range of thermodynamic variables, as discussed previously.

Trained solely on discrete EOS table entries using a mean-squared error loss, the auxiliary network implicitly learns a smooth interpolation manifold that represents the underlying thermodynamic relations of the EOS \cite{sitzmann2020implicit}. Unlike traditional piecewise linear or spline-based interpolation schemes, the periodic activations combined with Fourier feature embeddings allow the model to capture steep gradients and high-frequency structure without introducing spurious oscillations. Once trained, the auxiliary network serves as a global, differentiable surrogate of the stellar EOS, providing fast and consistent evaluations of $\rho(P,T,X,Z)$ during PINN training.

It should be highlighted that coordinate-based Multi-Layer Perceptrons (MLPs), which form the foundation of our PINN's architecture, inherently suffer from spectral bias; a phenomenon where the network rapidly learns low-frequency components but struggles to capture high-frequency details. This limitation is mathematically described by the Neural Tangent Kernel (NTK) framework \cite{jacot2018neural}, which demonstrates that the convergence rate of the training error is governed by the eigenvalues of the NTK matrix. For standard MLPs, the eigenvalues associated with high-frequency features are exceedingly small and decay rapidly. Consequently, the network converges exponentially slower on the steep gradients present in stellar profiles, such as the sharp physical transitions near the stellar core and surface, compared to the smooth, low-frequency global trends.

To overcome this spectral bias, we use the mentioned Random Fourier Features \cite{rahimi2007random} as the input layer of the architecture, which map the input coordinate (the initial mass $M_r$) into a high-dimensional space using the transformation $\mathrm{RFF}(M_r)=[\cos(2\pi \textbf{B}M_r),\sin(2\pi \textbf{B}M_r)]^\text{T}$. Here, the frequency matrix is sampled from a normal distribution, $\textbf{B}\sim \mathcal{N}(\mu,\,\sigma^{2})$. By processing the input through this RFF layer, we fundamentally alter the network's NTK into a stationary kernel, shifting its spectrum toward higher frequencies and enabling the model to accurately fit high-gradient regions by simply adjusting the variance $\sigma^2$. In our implementation, the $\sigma$ value is selected following \cite{shi2024adaptive} approach and a small set of stellar profiles calculated with \texttt{MESA}.

\section{Modeling energy sources}
\label{sec:energy-sources}
The luminosity equation (\eqref{eq:lum-r}) includes a net local energy source/sink term,
\begin{equation}
\varepsilon
=
\varepsilon_{\rm grav }
+
\varepsilon_{\rm nuc}
- \varepsilon_{\rm th},
\end{equation}
which accounts for nuclear energy generation, thermal neutrino losses, and energy exchange associated with gravitational contraction and expansion. In our formulation, the generation rate $\epsilon$ also depends exclusively on $(\rho, T, X_i)$. Given the variability and complexity of the function $\epsilon$, rather than computing it with another auxiliary network (as done before with the EOS for $\rho$ or for the opacity $\kappa$), it is now computed in our model following the classical approach described by MESA and briefly explained below. 

\subsection{Gravitational energy term}
A star may convert gravitational potential energy to thermal energy (and vice versa) through contraction or expansion. Using thermodynamic quantities supplied by the EoS (sec. \ref{sec:eos}), the gravitational term can be written as \cite{paxton2011modules}
\begin{equation}
\varepsilon_{\rm grav}
=
-\,T\,c_P
\left[
\left(1 - \chi_T \cdot \nabla_{\rm ad}\right)\frac{d\ln T}{dt}
-
\chi_\rho \cdot \nabla_{\rm ad}\,\frac{d\ln \rho}{dt}
\right],
\label{eq:eps-grav-cp}
\end{equation}
This formula is implemented in \texttt{MESA} using finite differences for the time derivative of the log stellar quantities $\rho$ and $T$. When a star is in hydrothermal equilibrium, the time derivatives are effectively null and the $\varepsilon_{\rm grav}$ term does not contribute as a source. 

\subsection{Nuclear energy generation}
Stars maintain their internal pressure and luminosity by converting nuclear binding energy into thermal energy through thermonuclear reactions. In stellar evolution theory, this conversion appears as a local source term in the energy (luminosity) equation. In this subsection, we explain how this source term is defined, how it is computed from nuclear reaction networks, and how it is incorporated into our physics-informed neural network (PINN) framework.

\paragraph{Energy source term in the stellar structure equations.}
The stellar structure equations describe how macroscopic quantities such as pressure, temperature, and luminosity vary inside a star. The luminosity $L(m)$ denotes the total energy per unit time flowing outward through a spherical surface that encloses a mass $m$. The derivative $dL/dm$ therefore represents the local rate at which energy is added to (or removed from) the stellar plasma per unit mass.

To close the system of stellar structure equations, one must specify the local energy source associated with nuclear reactions. This source is described by the nuclear energy generation rate
\begin{equation}
\varepsilon_{\rm nuc}(\rho, T, \{X_i\}),
\end{equation}
which gives the rate at which nuclear binding energy is released as or absorbed from the thermal energy of the plasma, per unit mass. Here $\rho$ is the local mass density, $T$ the local temperature, and $\{X_i\}$ the set of mass fractions of the nuclear species present, indexed by $i$.

The luminosity equation including nuclear energy generation can be written as
\begin{equation}
\frac{dL}{dm} = \varepsilon_{\rm nuc} - \varepsilon_{\nu,\rm nuc},
\end{equation}
where $\varepsilon_{\nu,\rm nuc}$ denotes the energy carried away by neutrinos produced directly in nuclear reactions. **The subscript $\nu$ indicates neutrino losses, while the subscript $\rm nuc$ specifies that these neutrinos originate from nuclear reactions rather than from thermal plasma processes.** Neutrinos weakly interact with the surrounding plasma and tend to escape freely from the stellar interior under normal conditions, sapping thermal energy from the star.

Nuclear energy generation is a strictly local quantity: it involves no spatial derivatives and depends only on the instantaneous thermodynamic state and chemical composition at a given mass coordinate, at a given time.

\paragraph{Reaction network as the origin of $\varepsilon_{\rm nuc}$.}
The nuclear energy generation rate is computed from a nuclear reaction network. The values in our MESA models are based on the \texttt{approx21.net} network. The network describes how nuclear species participate in reactions and how much energy is released or absorbed as a result.  Nuclear reaction data (thermal energy per nuclear reaction chain) in this case are taken from the Joint Institute for Nuclear Astrophysics (JINA) REACLIB database\cite{jina2007}. Nuclear species are labeled by an index $i$, and their abundances are expressed in terms of molar abundances
\begin{equation}
Y_i = \frac{X_i}{A_i},
\end{equation}
where $X_i$ is the mass fraction and $A_i$ the mass number of species $i$.

In a general time-dependent stellar evolution calculation, the molar abundances evolve according to
\begin{equation}
\frac{dY_i}{dt} = \sum_r \nu_{ir}\,\mathcal{R}_r(\rho,T,\{Y_j\}),
\end{equation}
where the sum runs over all nuclear reaction channels $r$, $\nu_{ir}$ is the net stoichiometric coefficient of species $i$ in reaction $r$, and $\mathcal{R}_r$ is the reaction progress rate.

In the present work, we do not solve these composition evolution equations dynamically. They are introduced only to define the nuclear energy generation rate under the assumption of a fixed chemical composition, appropriate for steady-state stellar structure.

\paragraph{Reaction progress rates.}
For a generic two-body thermonuclear reaction of the form $i + j \rightarrow k + \cdots$, the reaction progress rate is given by
\begin{equation}
\mathcal{R}_{ij} =
\frac{\rho N_A}{1+\delta_{ij}}\,
Y_i Y_j\,
\langle \sigma v \rangle_{ij}(T)\,
f^{\rm sc}_{ij}(\rho,T,\{Y\}),
\end{equation}
where $N_A$ is Avogadro’s number, $\delta_{ij}$ avoids double counting of identical reactants, $\langle \sigma v \rangle_{ij}$ is the Maxwellian-averaged thermonuclear reaction rate, and $f^{\rm sc}_{ij}$ is the electron-screening enhancement factor. Three-body reactions, such as the triple-$\alpha$ process, follow the same structure with an additional factor of $\rho N_A Y$.

Weak interactions and radioactive decays are treated as one-body processes,
\begin{equation}
\mathcal{R}_i^{\rm weak} = Y_i\,\lambda_i(\rho,T),
\end{equation}
where $\lambda_i$ denotes the corresponding weak interaction rate.

\paragraph{Definition of the nuclear energy generation rate.}
The nuclear energy generation rate is obtained by summing the energy released by all reaction channels,
\begin{equation}
\varepsilon_{\rm nuc}(\rho,T,\{Y\}) =
\sum_r Q_r\,\mathcal{R}_r(\rho,T,\{Y\}),
\end{equation}
where $Q_r$ is the thermal energy deposited in the plasma per reaction. The $Q$-values are derived from nuclear mass differences and include corrections for positron annihilation as well as the subtraction of energy carried away by neutrinos produced in weak reaction channels. As a result, $\varepsilon_{\rm nuc}$ represents the net local heating of the plasma due to nuclear reactions.

\paragraph{Physical regimes and stiffness.}
Different nuclear burning regimes dominate in different regions of the $(\rho,T)$ plane, such as the pp-chain and CNO cycle during hydrogen burning and the triple-$\alpha$ process during helium burning. The reaction rates depend extremely sensitively on temperature and span many orders of magnitude across stellar interiors. This stiffness, together with sharp regime transitions, makes nuclear energy generation challenging to be approximated by a neural network.

\paragraph{Treatment in the PINN framework.}
In our physics-informed neural network framework, nuclear energy generation is treated as an explicit physics operator. During training, the network-predicted thermodynamic state $(\rho,T,\{X_i\})$ is passed to a fixed nuclear microphysics module that evaluates $\varepsilon_{\rm nuc}$ pointwise. This design preserves physical fidelity and allows the neural network to focus on solving the global stellar structure equations rather than learning highly stiff microphysical processes.

\subsection{Thermal losses}
In addition to energy conversion through nuclear reactions, stars lose energy through neutrino emission processes that are not directly tied to nuclear burning. These processes remove energy from the stellar plasma and act as local sink terms in the energy equation. In this subsection, we describe the physical origin of these thermal losses and how they are incorporated into the stellar structure equations and the PINN framework.

\paragraph{Thermal neutrino losses in the energy equation.}
Thermal neutrino losses are described by a specific energy loss rate
\begin{equation}
\varepsilon_{\nu,\rm th}(\rho, T, \{X_i\}),
\end{equation}
which depends on the local density $\rho$, temperature $T$, and chemical composition. These losses enter the luminosity equation as
\begin{equation}
\frac{dL}{dm} = \varepsilon_{\rm nuc} - \varepsilon_{\nu,\rm th}.
\end{equation}
Unlike photons, which transport energy outward through radiative diffusion, neutrinos escape freely from stellar interiors under normal conditions, in which case they remove energy locally.

\paragraph{Neutrino emission mechanisms.}
Several physical processes contribute to thermal neutrino emission, including electron--positron pair annihilation, plasmon decay, photoneutrino production, and bremsstrahlung during electron--ion scattering. Each mechanism dominates in a different region of the $(\rho,T)$ parameter space, but all play the same conceptual role as local energy sinks, producing neutrinos that tend to remove heat.

\paragraph{Regime dependence.}
Thermal neutrino losses are negligible in low-mass main-sequence stars but become increasingly important in massive stars and during advanced evolutionary stages, where high temperatures and densities enhance neutrino emission rates. The relative importance of different emission mechanisms varies smoothly but nonlinearly across the stellar interior.

\paragraph{Mathematical formulation.}
The total thermal neutrino loss rate can be written as a sum over contributing processes,
\begin{equation}
\varepsilon_{\nu,\rm th}(\rho,T) =
\sum_k \varepsilon_{\nu}^{(k)}(\rho,T),
\end{equation}
where the index $k$ labels the individual emission mechanisms. In practice, each term is evaluated using analytic fits or tabulated expressions that ensure numerical stability and continuity across different physical regimes.

\paragraph{Coupling to the equation of state.}
Thermal neutrino emission depends sensitively on the thermodynamic state of the plasma, including electron degeneracy and relativistic effects. For consistency, the evaluation of $\varepsilon_{\nu,\rm th}$ must be compatible with the adopted equation of state to avoid double counting of energy losses.

\paragraph{Treatment in the PINN framework.}
As with nuclear energy generation, thermal neutrino losses are treated as fixed physics operators in the PINN framework. During training, the neural network provides the local thermodynamic state $(\rho, T, \{X_i\})$, which is passed to a nuclear and neutrino microphysics module taken directly from MESA and not learned. This module evaluates the corresponding neutrino loss rate pointwise using established microphysical prescriptions. This separation preserves physical fidelity and ensures that the neural network focuses on solving the global stellar structure equations rather than attempting to learn highly nonlinear and regime-dependent microphysical loss processes.

\bibliography{sample}